\documentclass[11pt,a4paper]{article}
\pdfoutput=1
\usepackage{jinstpub}

\RequirePackage{graphicx}
\RequirePackage{flushend}
\RequirePackage[numbers,sort&compress]{natbib}

\usepackage{amsfonts}
\usepackage{amssymb}
\usepackage{amsmath}
\usepackage{wasysym}
\usepackage{multirow}
\usepackage{subfigure}
\usepackage{siunitx}
\usepackage{color}
\usepackage[switch]{lineno}

\RequirePackage[capitalize]{cleveref}
\crefname{figure}{fig.}{figs.}
\crefname{table}{table}{tables.}
\crefname{equation}{eqn.}{eqns.}
\crefname{section}{sec.}{secs.}

\begin{document}

\title{Cryogenic Characterization of 180nm CMOS Technology at 100mK}

\author[a,b]{R.~G.~Huang}
\author[b]{D.~Gnani}
\author[b]{C.~Grace}
\author[a,b]{Yu.~G.~Kolomensky}
\author[a,b]{Y.~Mei}
\author[b]{A.~Papadopoulou}

\affiliation[a]{University of California, Berkeley, CA 94720-7300, USA}
\affiliation[b]{Lawrence Berkeley National Laboratory, CA 94720-8153, USA}

\emailAdd{roger\_huang@berkeley.edu}

\abstract{Conventional CMOS technology operated in cryogenic conditions has recently attracted interest for its use in low-noise electronics. We present one of the first characterizations of \SI{180}{nm} CMOS technology at a temperature of \SI{100}{mK}, extracting I/V characteristics, threshold voltages, and transconductance values, as well as observing their temperature dependence. We find that CMOS devices remain fully operational down to these temperatures, although we observe hysteresis effects in some devices. The measurements described in this paper can be used to inform the future design of CMOS devices intended to be operated in this deep cryogenic regime.}

\date{Received: date / Accepted: date}

\maketitle

\section{Introduction}
\label{sec:Introduction}

Microelectronics manufactured using conventional CMOS processes but operated at cryogenic temperatures of \SI{4}{K} and below have recently attracted interest in quantum computing for their use as precision controllers and low noise amplifiers \cite{CMOSQuantumComputing,CMOSQuantumComputing2}. This method of incorporating electronics directly into the cryogenic environment instead of operating them at room temperature can offer similar advantages in experiments like CUORE (Cryogenic Underground Observatory for Rare Events), which uses a cryogenic bolometric approach to search for neutrinoless double beta decay. CUORE uses Neutron Transmutation Doped (NTD) thermistors on TeO$_2$ crystals to sense temperature changes induced by physical energy deposits. At the moment, all the CUORE electronics, including those for biasing the NTDs, amplifying the signals, and performing readout, are operated at room temperature \cite{CUOREElectronics}. The future CUORE Upgrade with Particle ID (CUPID) plans to take advantage of the general cryogenic infrastructure developed for CUORE, but an upgrade to its electronics infrastructure is under consideration \cite{CUPIDInterestGroup:2019inu}.  CMOS microelectronics engineered to be operated at or below \SI{4}{K} offer an alternative approach to signal preamplification in CUPID, allowing a reduction in electronic noise and a possibility of introducing a modest channel multiplexing factor.

To date, there are very few measurements of CMOS device properties at sub-Kelvin temperatures, which will have to be understood if we wish to consider using them to construct amplifiers and multiplexers near the base operating temperature of CUPID.  In this paper we present one of the first characterizations of \SI{180}{nm} CMOS technology down to \SI{100}{mK}, which will be used to inform the design of these devices.

\section{Setup}
\label{sec:Setup}

\begin{figure}
\centering
\includegraphics[width=0.85\textwidth]{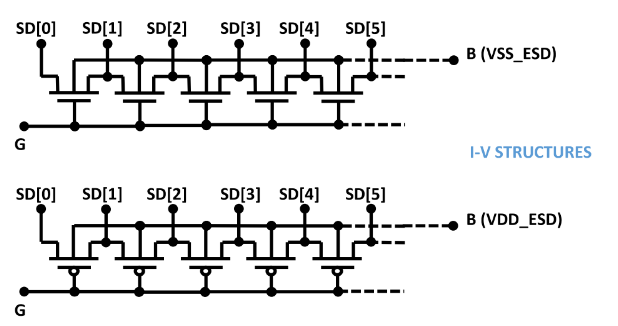}
\caption{\label{fig:IVStructure} PMOS and NMOS structures on the test chip for I/V characterization measurements.  Each labeled point corresponds to a $\SI{72}{\micro m}\times\SI{63}{\micro m}$ sized pad that can be wirebonded from the outside.}
\end{figure}

We conduct our characterization measurements with a TSMC \SI{180}{nm} CMOS test chip containing multiple arrays of PMOS and NMOS devices of a variety of thresholds with the widths and lengths varying from \SI{180}{nm}  to \SI{10}{\micro m}.  The MOSFET structures for quasi-static voltage-current (I/V) scan characterization are shown in Fig.~\ref{fig:IVStructure}.  We selectively wirebond the devices to be measured out of the array.  The chip is placed at the mixing chamber stage of an Oxford Triton-400 dilution refrigerator, kept at \SI{100}{mK} for these measurements.  For voltage biasing and current readout, the MOSFETs are wired out to a Keithley 2450 SMU and a Keithley 6517B electrometer outside the cryostat.  Since the dilution refrigerator has limited cooling power, the device power consumption was kept to a minimum by limiting the input voltage, the current range, and the amount of time they were kept on.

\section{Results}
\label{sec:Results}
\subsection{Characterization}

\begin{figure}
\centering
\includegraphics[width=0.48\textwidth]{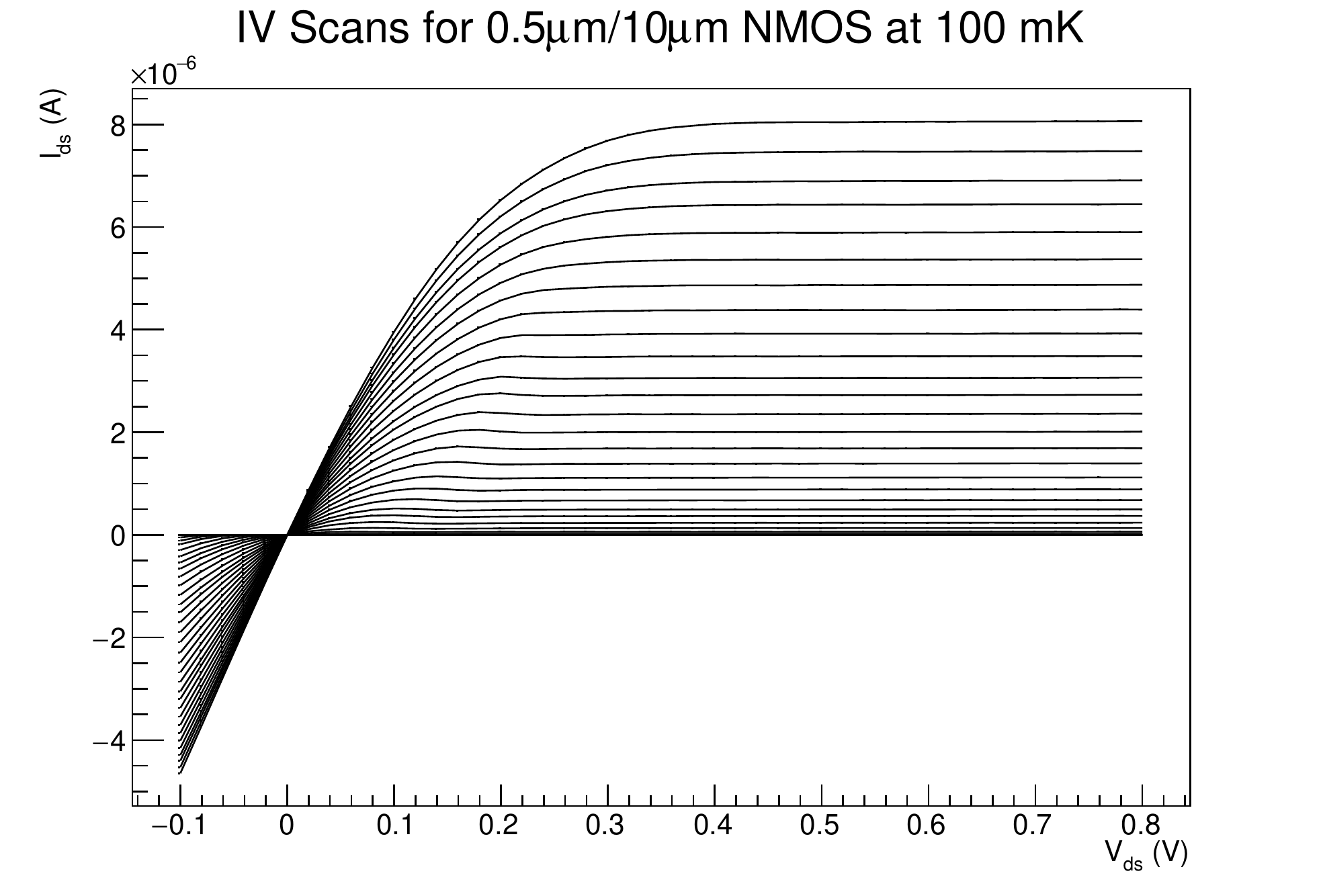}
\includegraphics[width=0.48\textwidth]{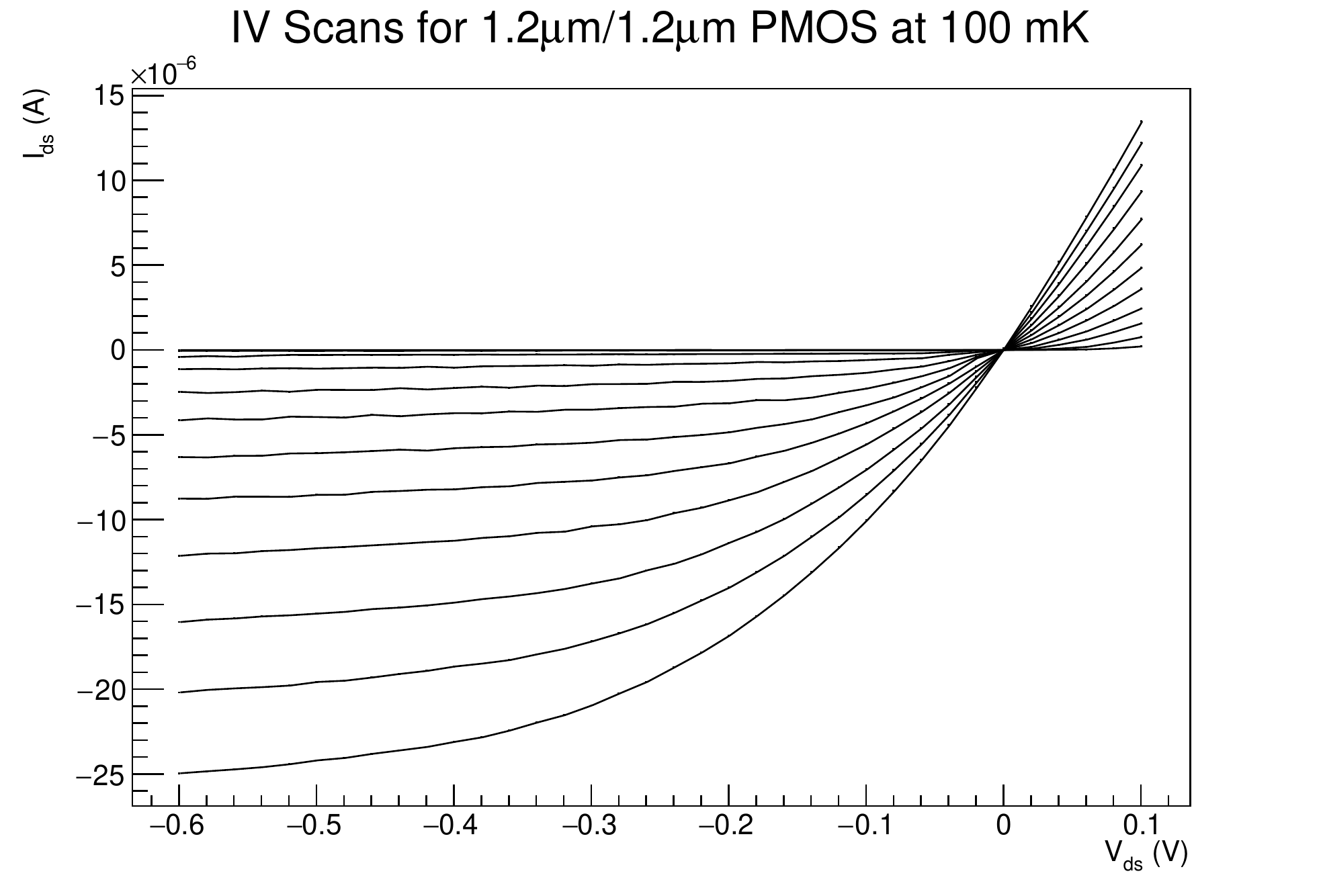}
\caption{\label{fig:IVCurves} $I_d$ vs.\ $V_d$ curves taken at a temperature of \SI{100}{mK}.  Each curve in a plot corresponds to a different gate voltage $V_g$.  Left: $W/L=0.5/10[\si{\micro m}]$ NMOS, with $V_g$ scanned from 0.4 to 1.2 V in 0.02 V increments.  Right: $W/L=1.2/1.2[\si{\micro m}]$ PMOS, with $V_g$ scanned from 0.2 to 0.8 V in 0.05 V increments.}
\end{figure}

\begin{figure}
\centering
\includegraphics[width=0.48\textwidth]{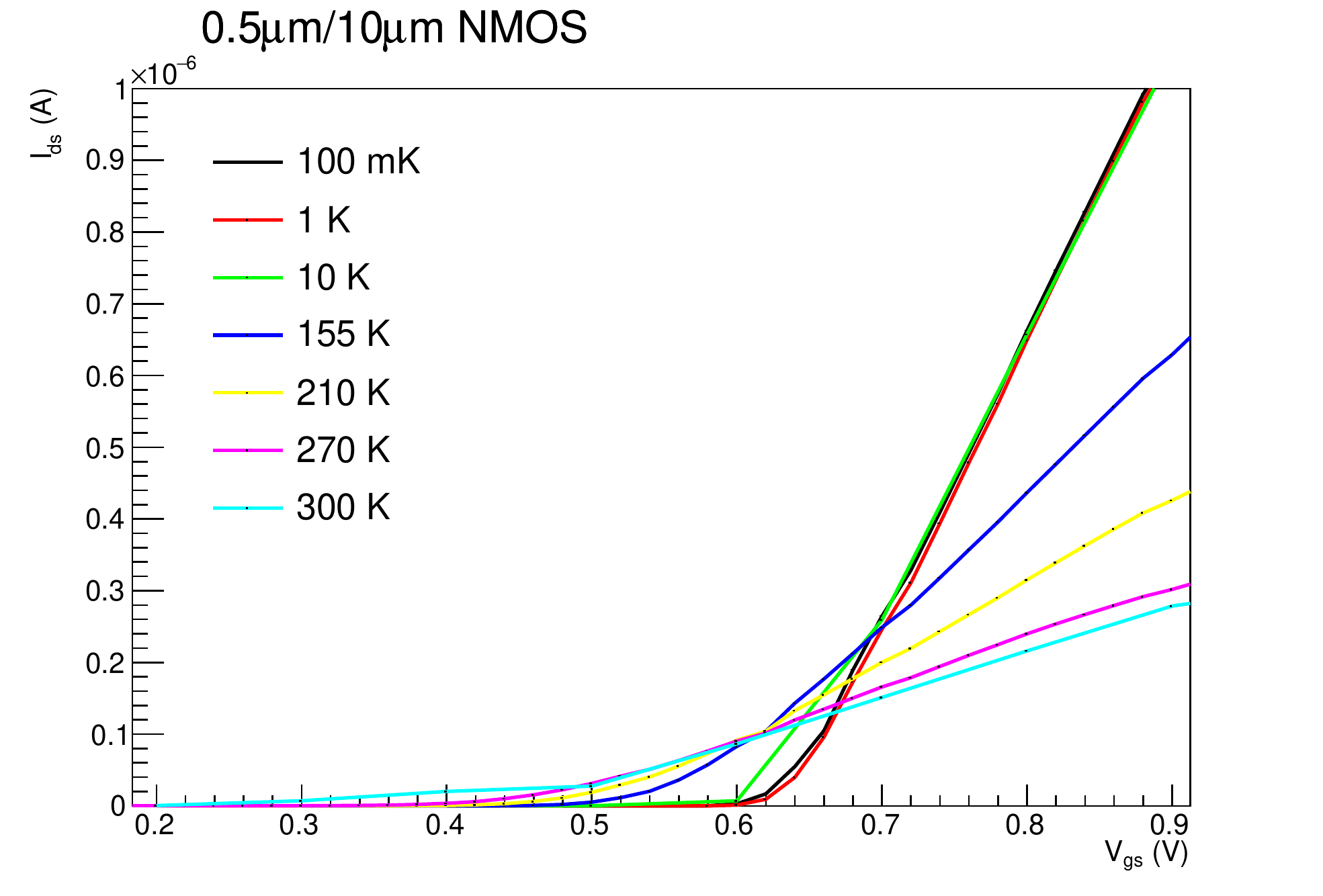}
\includegraphics[width=0.48\textwidth]{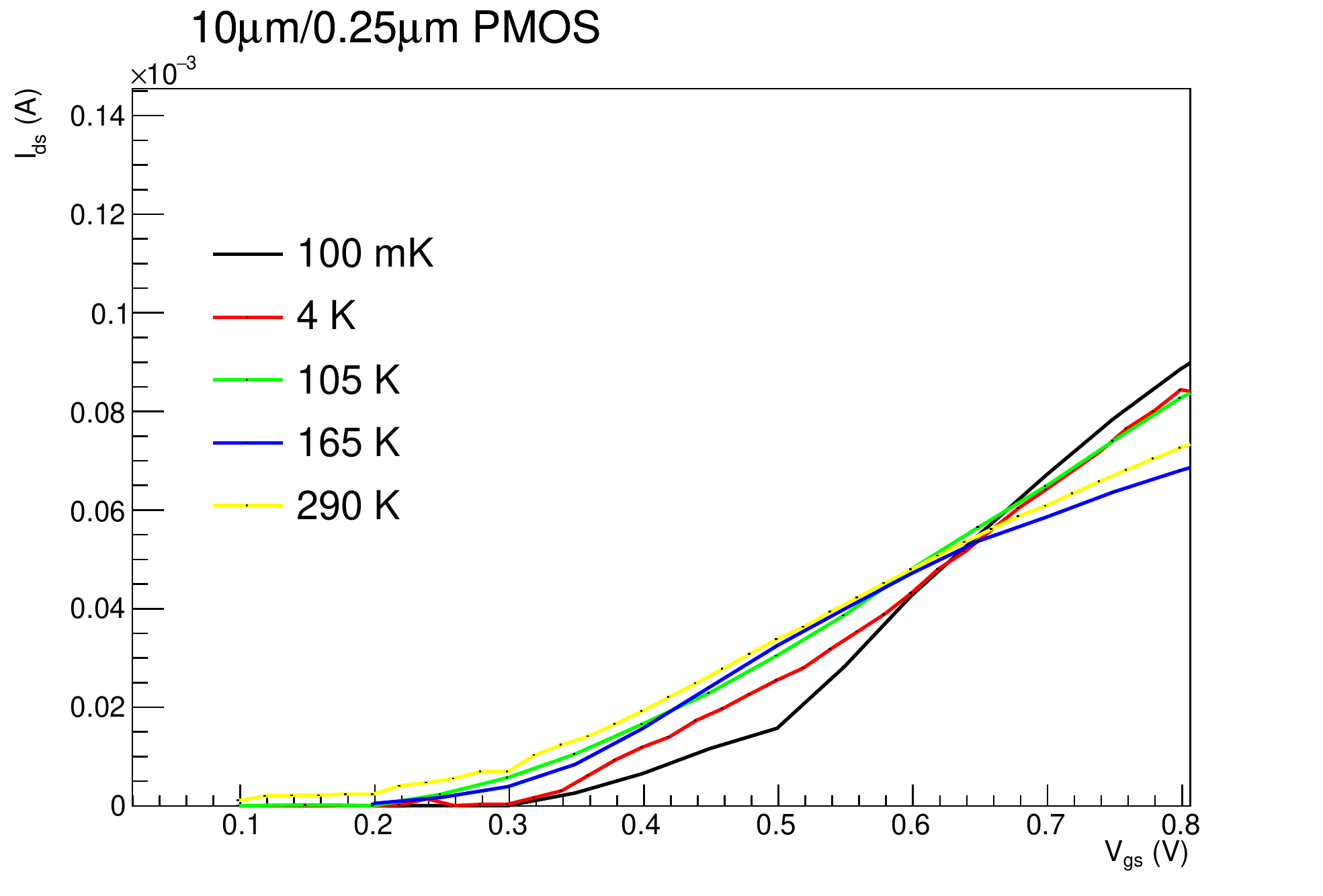} \\
\includegraphics[width=0.48\textwidth]{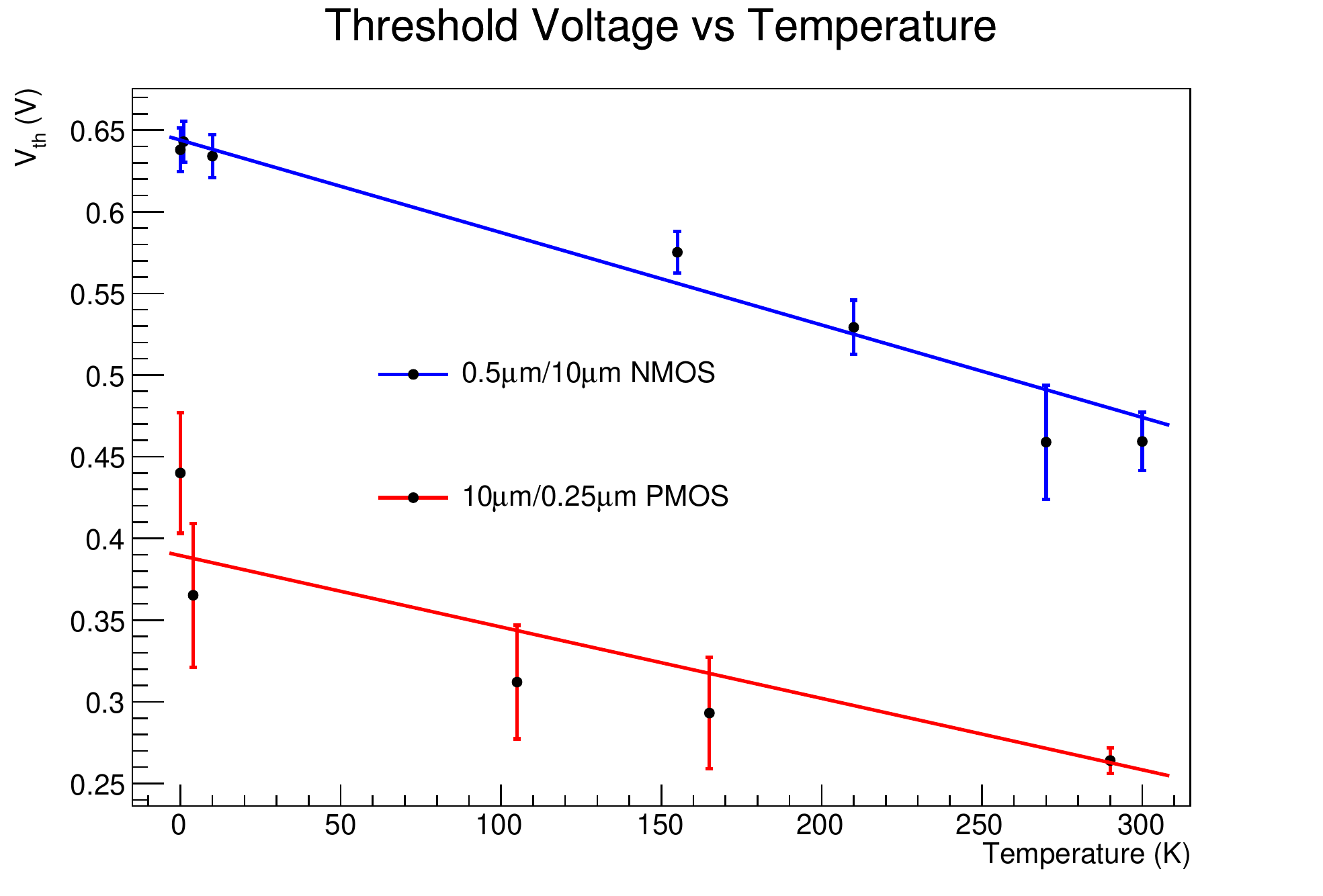}
\includegraphics[width=0.48\textwidth]{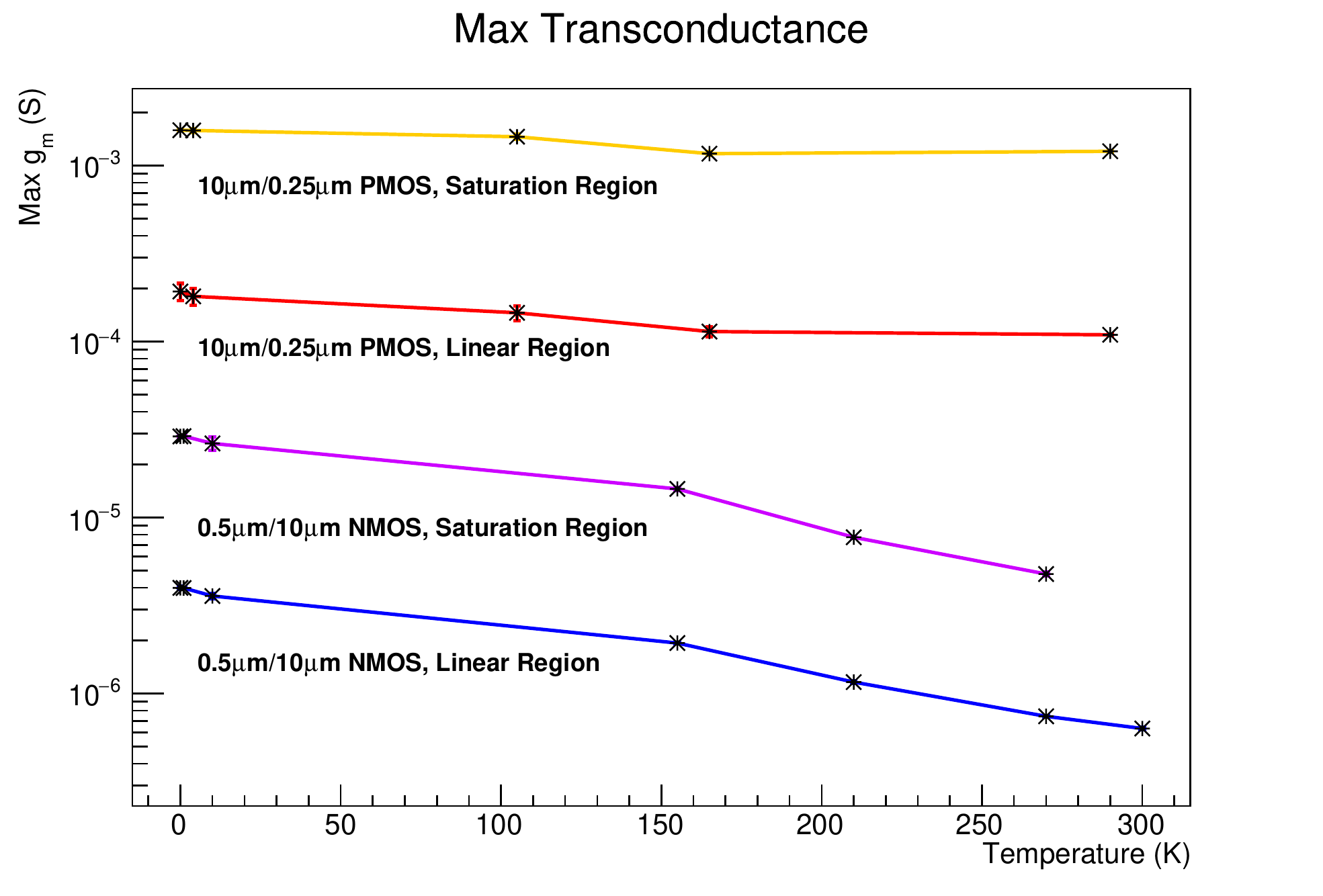}
\caption{\label{fig:Parameters} Top left: $I_d$/$V_g$ plot for a $W/L=0.5/10[\si{\micro m}]$ NMOS, evaluated at $V_d=0.02$ V.  Top right: $I_d$/$V_g$ plot for a $W/L=10/0.25[\si{\micro m}]$ PMOS, evaluated at $V_d=0.02$ V.  Bottom left: Extracted threshold voltages for each device as a function of temperature.  The lines are fits for $V_{th}$ as a linear function of temperature, with $\chi^2/DoF=4.1/7$ for the NMOS and $\chi^2/DoF=3.5/5$ for the PMOS.  Bottom right: The maximum transconductance in the scan range for each device, evaluated in both the linear and saturation regions, as a function of temperature.}
\end{figure}

We perform standard I/V scans for an assortment of PMOS and NMOS devices on the test chip.  All MOSFETs tested remain operational down to temperatures of $<\SI{100}{mK}$; examples of I/V scan curves are shown in Fig.~\ref{fig:IVCurves}.  From these measurements, we extract the transconductance and the threshold voltage.  Their temperature dependence is shown in Fig.~\ref{fig:Parameters}.  The threshold voltage is obtained by plotting the drain current $I_d$ against the gate voltage $V_g$, extrapolating from the linear region of the $I_d$-$V_g$ curve, and then finding the intersection with the $V_g$ axis \cite{ThresholdVoltageMethods}.  We observe that the threshold voltage increases as the temperature goes down; general models of MOSFET behavior predict that threshold voltage should scale as $V_{th} \propto (V_{th,0}-T)$, and our data suggests no significant deviations from this behavior all the way down to \SI{100}{mK}. Uncertainties in the threshold voltages are primarily due to uncertainties in the linear extrapolation of the $I_d$-$V_g$ curves.

We select a simple square-law MOSFET model in order to obtain an approximation of transconductance in both the linear region and the saturation region.  The drain current in each of these regions is given by
\[I_D = 2k\left[\left(V_g - V_T\right)V_d - 0.5V_d^2\right] \text{ (linear region)}\]
\[I_D = k\left(V_g-V_T\right)^2 \text{ (saturation region)}\]
where $k$ includes details from the characteristic feature size and manufacturing process of the MOSFET and generally has a temperature dependence of $k \propto T^{-3\text{/}2}$ due to carrier mobility effects.  For both of these regions, this gives an expected $g_m \propto T^{-3\text{/}2}$ relationship.  However, while we observe that $g_m$ does increase at lower temperatures, this relation does not seem to hold for the entire temperature range. This indicates that this basic model likely becomes invalid at sufficiently low temperatures, where other effects such as carrier freeze-out may become relevant.

\subsection{Hysteresis Effects}

\begin{figure}
\centering
\includegraphics[width=0.48\textwidth]{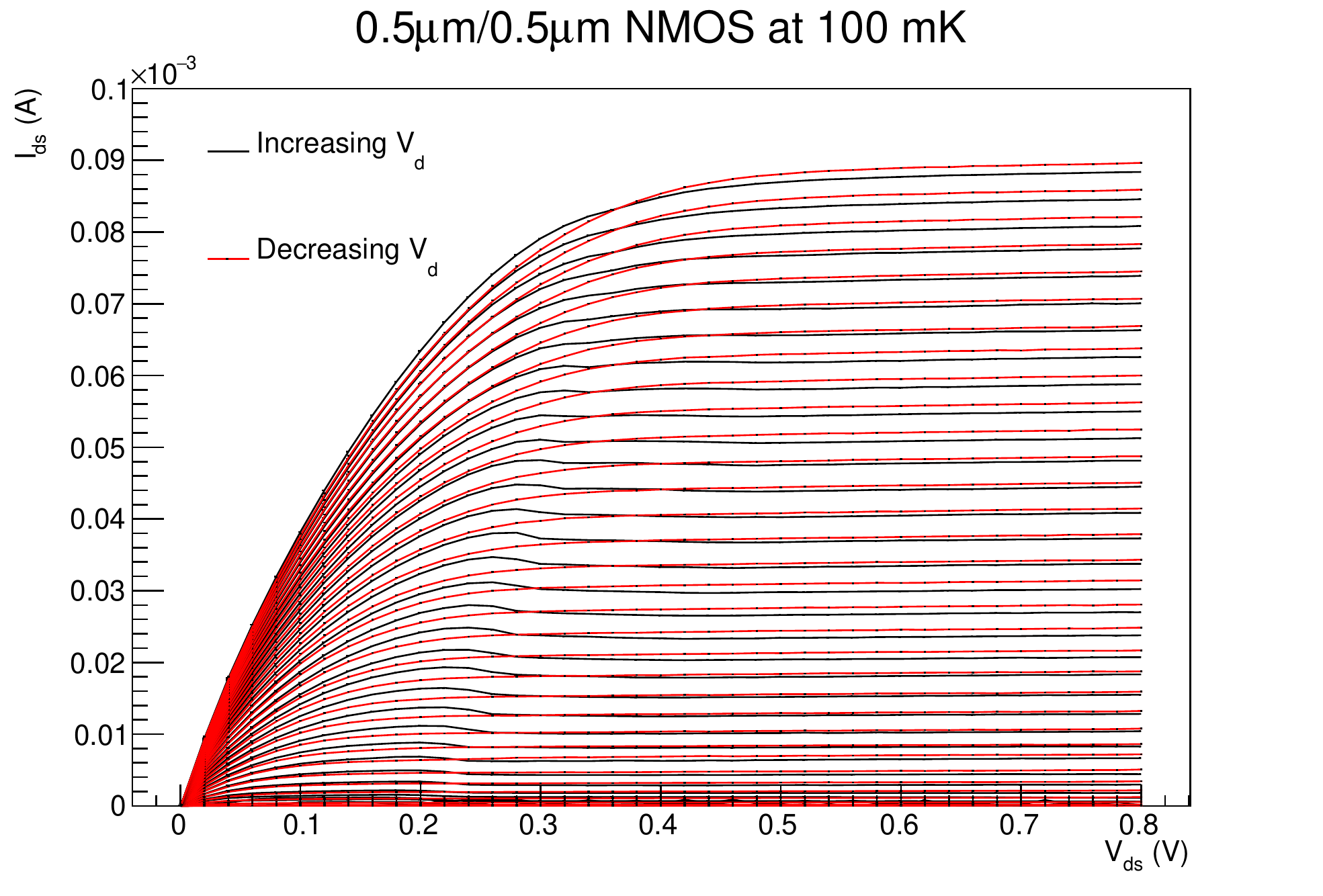}
\includegraphics[width=0.48\textwidth]{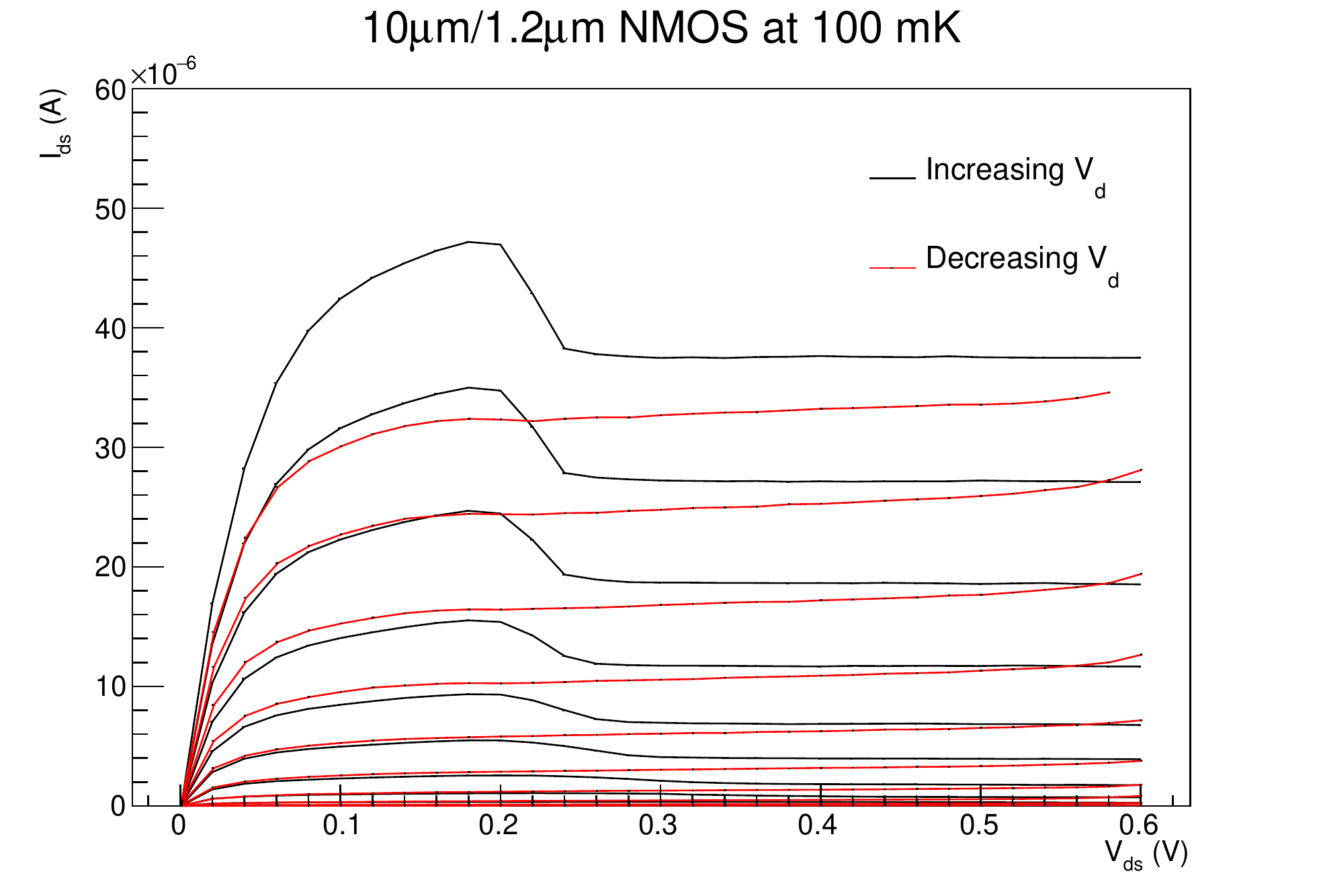} \\
\includegraphics[width=0.48\textwidth]{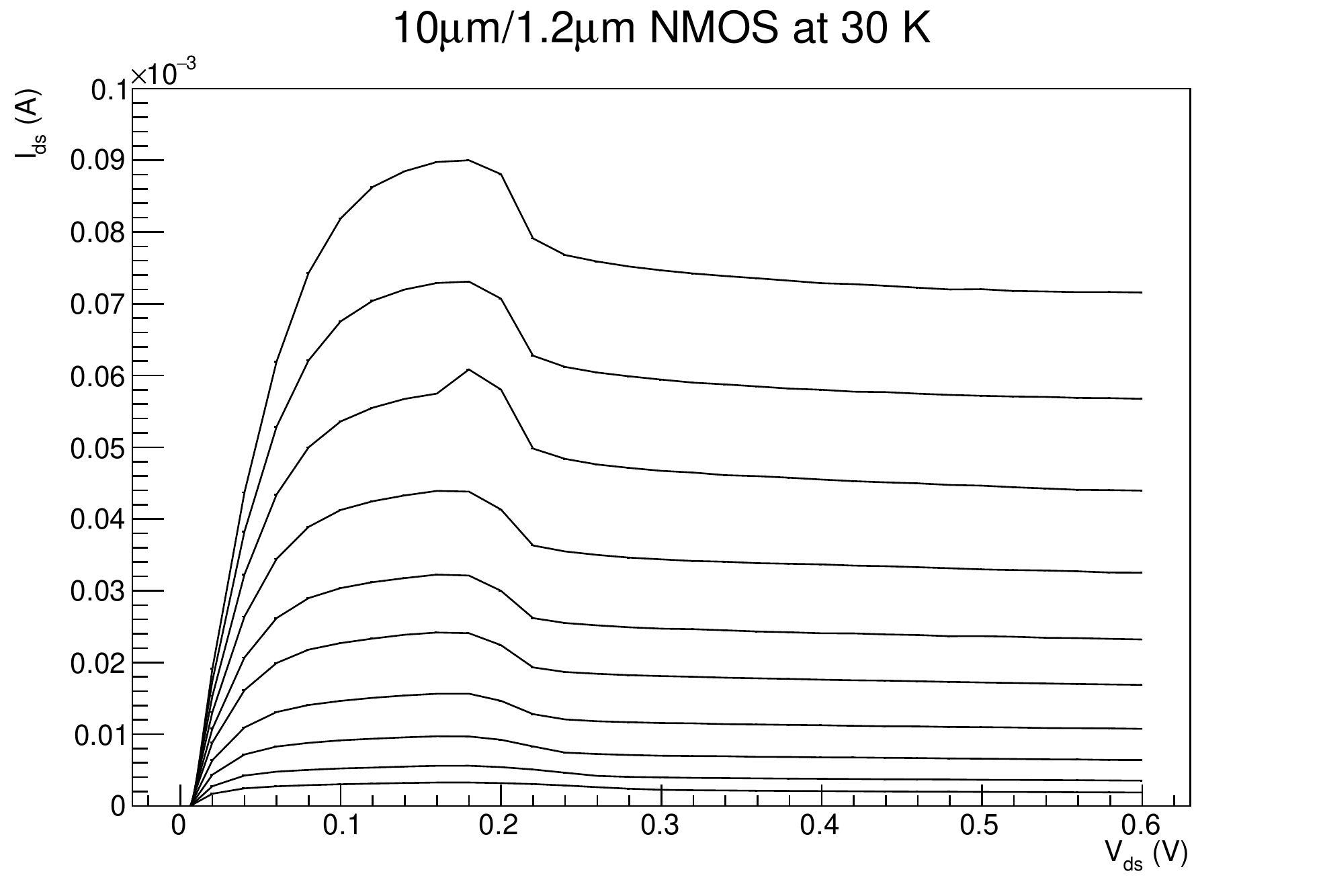}
\includegraphics[width=0.48\textwidth]{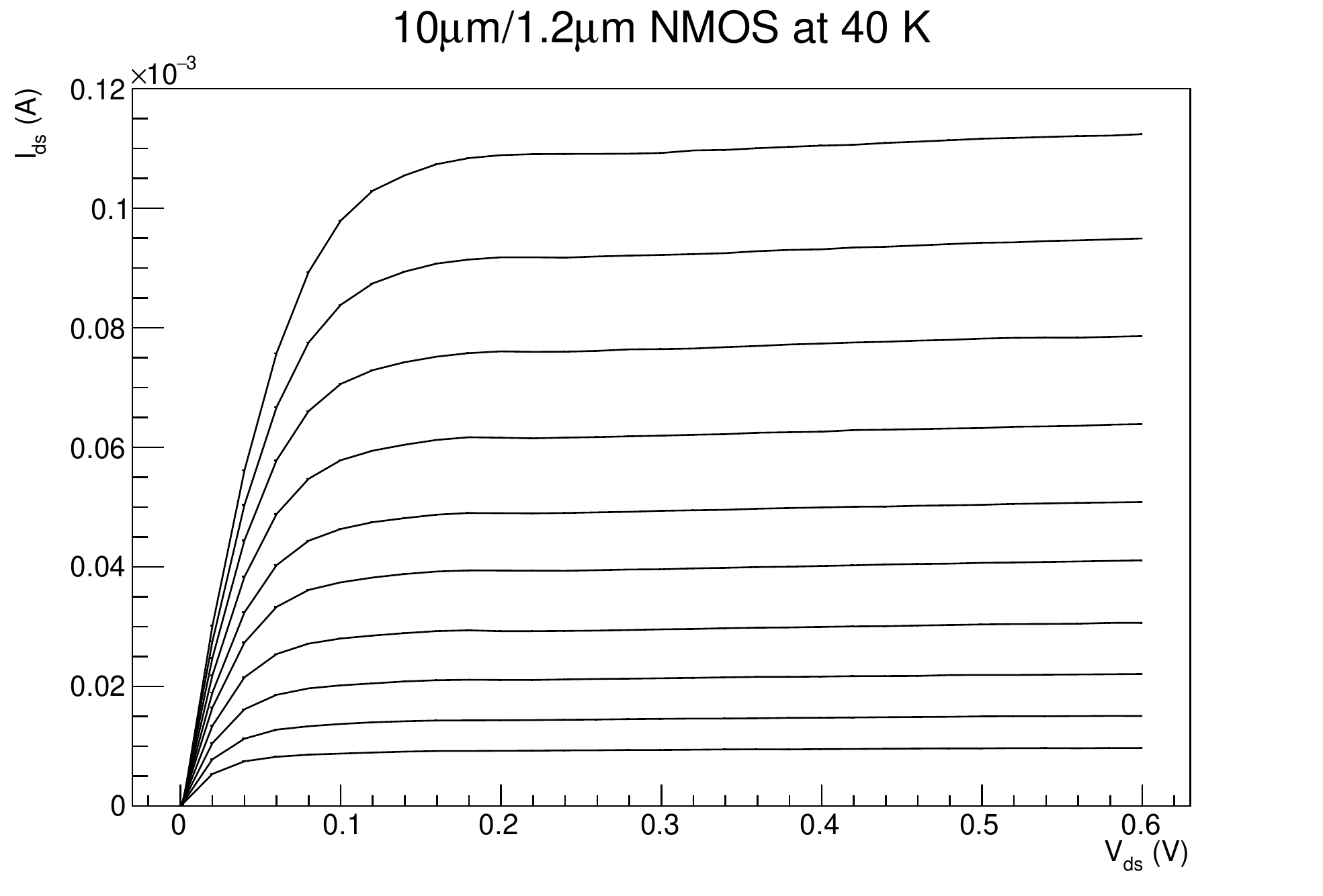}
\caption{\label{fig:IVCurveKink} Characteristic I/V curves with $V_g$ scanned in 0.02 V increments, showing the presence of a temperature-dependent hysteresis.  Top left: $W/L=0.5/0.5[\si{\micro m}]$ NMOS at \SI{100}{mK}, for both increasing and decreasing scans of $V_d$, where it can be seen that the kinks are only visible when $V_d$ is increased.  Top right: $W/L=10/1.2[\si{\micro m}]$ NMOS at \SI{100}{mK}, where the same kind of behavior is observed.  Bottom left: $W/L=10/1.2[\si{\micro m}]$ NMOS at \SI{30}{K}, where the kinks are still present but less prominent.  Bottom right: $W/L=10/1.2[\si{\micro m}]$ at \SI{40}{K}, where the kinks have disappeared.}
\end{figure}

Some of the MOSFETs exhibit a hysteresis effect when operated at low enough temperatures, with a kink appearing in the characteristic I/V curves when scanning from low $V_d$ to high $V_d$, but not when going from high $V_d$ to low $V_d$, as shown in Fig.~\ref{fig:IVCurveKink}.  This effect is observed in different sized devices to different degrees, and the effect disappears once the temperature of the devices is high enough.  The temperature dependence of this feature indicates it is likely related to charge carrier freeze-out in the MOSFETs.  A similar kink in the I/V curves at cryogenic temperatures has previously been observed in \cite{MOSFETHysteresis}, where it was explained by the formation of a forced depletion layer near the pinchoff region of the MOSFET.  Taking NMOS for example, in this model, $V_d$ is high enough to cause impact ionization in the pinchoff region of the transistor as it starts to form.  The electrons freed by this process then join the drain current, while the holes are pulled deeper into the substrate, enjoying the high carrier mobility induced by the low temperatures.  The freed electrons from this initial forced formation of the depletion layer cause an increase in the drain current, but after the layer is formed it allows charge carriers coming from the inversion layer to pass through it normally without further interaction.  This effect is only present at temperatures where carrier freeze-out is significant, as otherwise the depletion layer in the pinchoff region can form normally from the movement of thermally excited charge carriers.  The kink also does not occur when scanning from high $V_d$ to low $V_d$, as the extra current caused by the formation of the depletion layer does not change the monotonically decreasing drain current as drain voltage is decreased.

This explanation matches several of the characteristics of the hysteresis effect that we have observed.  The magnitude of the kink in the I/V curves is smaller for NMOS with smaller channel sizes, as seen in the top figures of Fig.~\ref{fig:IVCurveKink}, which matches the expectation of a smaller current surge from the formation of a smaller depletion layer.  We observe that the effect gradually decreases in size as temperature increases until it disappears around \SI{40}{K}, which is where carrier freeze-out should start becoming insignificant in our silicon-based substrates.  We also did not observe this effect in our PMOS devices (see Fig.~\ref{fig:IVCurves}), which matches the model's expectation that forced depletion layer formation is suppressed in PMOS due to the substantially lower substrate currents.

\subsection{Simulation}

\begin{figure}
\centering
\includegraphics[width=0.48\textwidth]{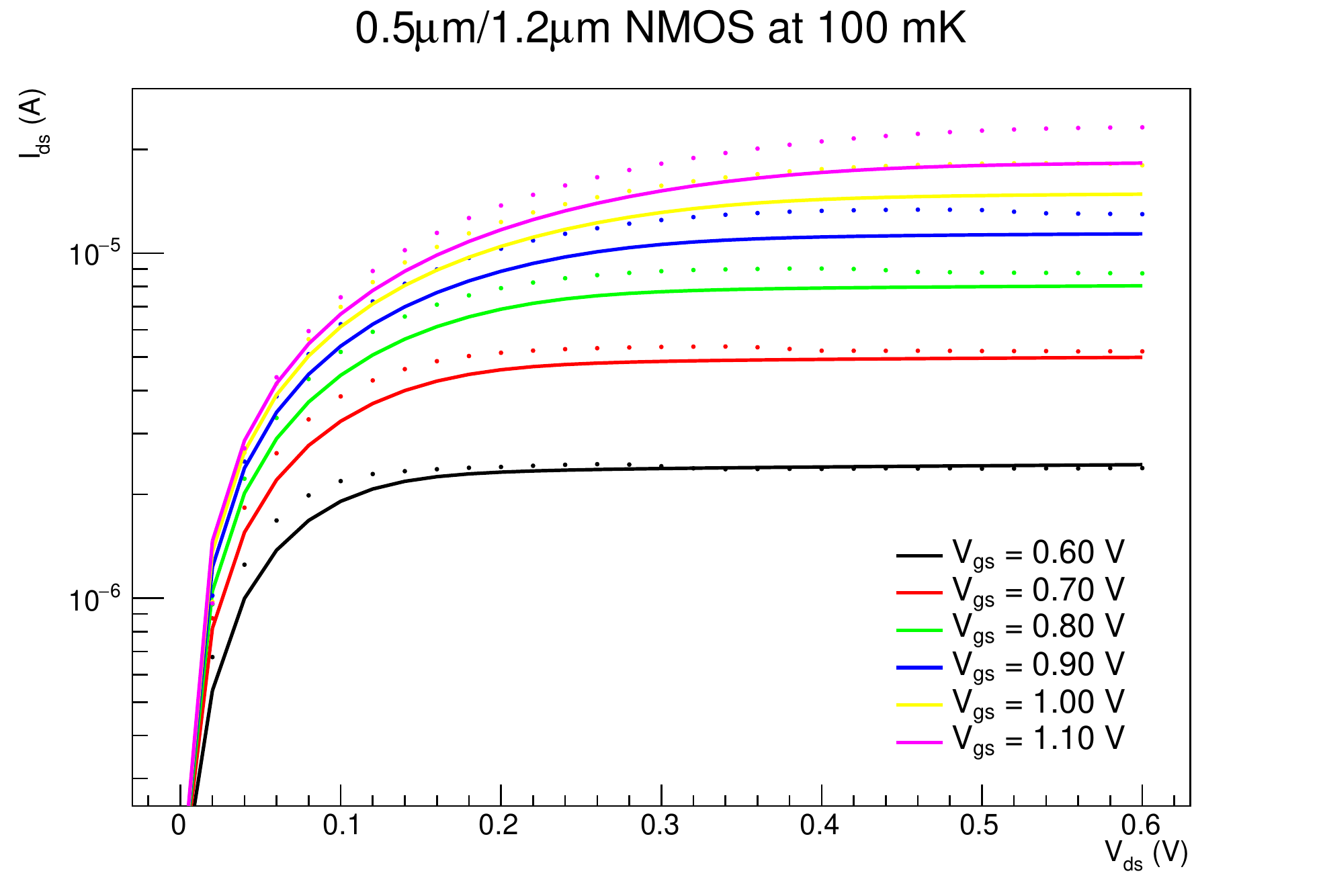}
\includegraphics[width=0.48\textwidth]{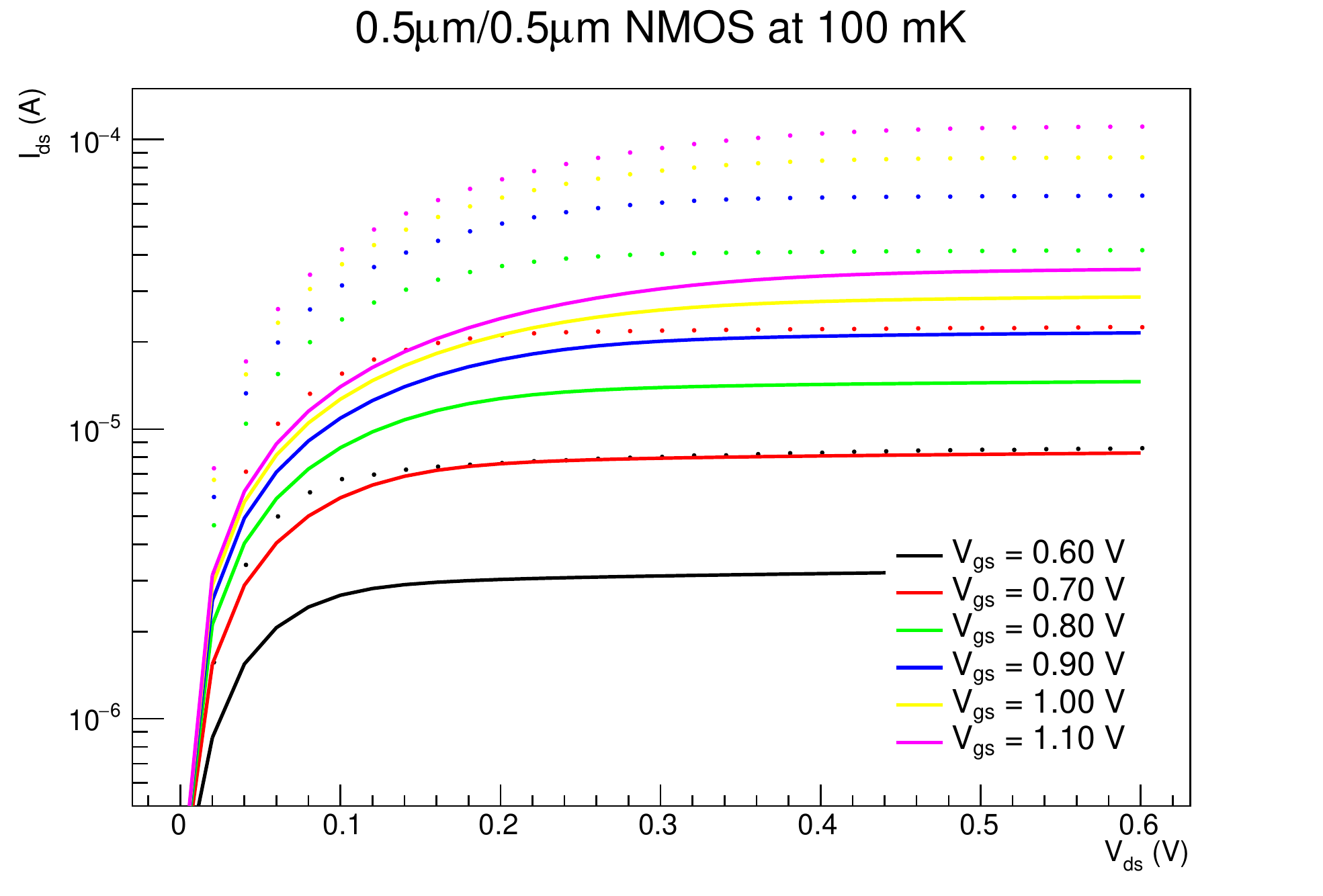}
\caption{\label{fig:BSIMFits} Simulated $I_d-V_d$ curves from a BSIM3 model tuned to measured NMOS behavior at 100 mK, with the simulated response drawn as lines and measured response shown as points. Left: $0.5/1.2[\si{\micro m}]$ NMOS that the model parameters are tuned to. Right: $0.5/0.5[\si{\micro m}]$ NMOS using the model parameters obtained from the left. The model can be tuned to reasonable agreement with the size shown on the left, but the result does not accurately predict the behavior of a different sized device shown on the right.}
\end{figure}

We apply a basic BSIM3 model to try to replicate the behavior of our devices at cryogenic temperatures, tuning a limited set of the model parameters to fit the measurements \cite{BSIMManual}. The BSIM3 model is not designed to work below around 220 K, but previous work has shown it is possible to phenomenologically tune the model parameters to work down to 77 K \cite{BSIMCryogenicModel,BSIMCryogenicModel2}, and possibly even down to 4 K \cite{CryoCMOSCharacterization}. We tune the $V_{th0}, K_1, U_A,$ and $U_B$ parameters using measurements from one device size and then apply the resulting model to other sizes for comparison. An example is shown in Fig. \ref{fig:BSIMFits}, where the BSIM parameters are tuned to a $0.5/1.2[\si{\micro m}]$ NMOS at 100 mK but it can be seen that the resulting model does not accurately predict the I/V characteristics of a $0.5/0.5[\si{\micro m}]$ NMOS at the same temperature. However, the simulated I/V curves obtained from the BSIM3 model work well for all device sizes at room temperature as expected, so this suggests that the size-dependence of the I/V characteristics changes substantially at very low temperatures in a manner that we have not captured. Further tuning of this model will be possible with measurements on a larger range of device sizes.  

\section{Conclusions}
\label{sec:Conclusion}

In this work, we have demonstrated successful operation of \SI{180}{nm} CMOS technology down to temperatures of \SI{100}{mK} and performed one of the first characterizations of its properties in this temperature range.  We find that CMOS behavior at this temperature is qualitatively similar to the behavior at \SI{4}{K} observed in previous works, which is a first step towards establishing that their usage as signal amplifiers or multiplexers can be extended down to temperatures as low as \SI{100}{mK}. Such amplifiers will likely use MOSFETs operated in the weak inversion region in order to satisfy the power budget allowed by a dilution refrigerator at these temperatures. While we have observed significant hysteresis effects in some NMOS devices, this can be accounted for in amplifier designs to avoid any undesirable effects on amplifier performance. Future extensions of this work will involve measurements with a larger range of device sizes to fine-tune our models of cryogenic behavior, as well as testing CMOS-based signal amplifiers in the 10 to \SI{100}{mK} range, with the goal of minimizing noise and optimizing performance within the power dissipation constraints imposed by a dilution refrigerator.

\section{Acknowledgments}
We would like to thank the members of the Kolomensky group for support in arranging the conditions for our cryogenic measurements. This work was supported by the U.S.\ Department of Energy under Contract No.~DE-AC02-05CH11231 and by the DOE Office of Science, Office of Nuclear Physics under Contract Nos. DE-FG02-08ER41551 and DE-SC0017617.

\bibliographystyle{JHEP}
\bibliography{Bibliography}

\end{document}